\documentclass[12pt]{article}
\usepackage{amsmath,amssymb}
\usepackage{amsthm}
\usepackage{epsfig}
\usepackage{latexsym}

\renewcommand{\thefootnote}{\fnsymbol{footnote}}
\newcounter{aff}
\renewcommand{\theaff}{\fnsymbol{aff}}
\newcommand{\affiliation}[1]{
\setcounter{aff}{#1} $\rule{0em}{1.2ex}^\theaff\hspace{-.4em}$}

\def\cN{\mathcal{N}}

\def\mint{\int_{-\infty}^\infty\!\cdots\!\int_{-\infty}^\infty}
\def\l{\ell}

\newcommand{\be}{\begin{equation}}
\newcommand{\ee}{\end{equation}}
\newcommand{\ba}{\begin{aligned}}
\newcommand{\ea}{\end{aligned}}

\def\({\left(}
\def\){\right)}

\DeclareMathOperator{\real}{Re}
\DeclareMathOperator{\im}{Im}

\begin{document}
\begin{titlepage}
\hfill\hfill
\begin{minipage}{1.2in}
RUP-20-24
\end{minipage}

\bigskip\bigskip\bigskip\bigskip
\begin{center}
{\LARGE\bf An Alternative to the Teukolsky Equation}

\bigskip\bigskip
{\large Yasuyuki Hatsuda\footnote[1]{\tt yhatsuda@rikkyo.ac.jp}
}\\
\bigskip\bigskip
\affiliation{1}
{\normalsize\it Department of Physics, Rikkyo University, Toshima, Tokyo 171-8501, Japan} 
\end{center}

\begin{abstract}
We conjecture a new ordinary differential equation exactly isospectral to the radial component of the homogeneous Teukolsky equation.
We find this novel relation by a hidden symmetry implied from a four-dimensional $\mathcal{N}=2$ supersymmetric quantum chromodynamics.
Our proposal is powerful both in analytical and in numerical studies.
As an application, we derive high-order perturbative series of quasinormal mode frequencies in the slowly rotating limit. We also test our result numerically by comparing it with a known technique.
\end{abstract}

\end{titlepage}

\renewcommand{\thefootnote}{\arabic{footnote}}
\setcounter{footnote}{0}
\setcounter{section}{0}

\section{Introduction}
The Kerr geometry is widely thought to be a good approximation to astrophysical black holes. They are significant both in astrophysics and in theoretical physics. In 1972, Teukolsky derived a master wave equation in linear perturbation theory of the Kerr black holes \cite{teukolsky1972}.
Now, Teukolsky's equation is a fundamental tool to investigate the rotating black holes.
Its significance is increasing more and more after the observation of gravitational waves \cite{abbott2016}.

The most remarkable point on the Teukolsky equation is that it is separable to the radial and the angular components even though the background is not spherically symmetric nor static.
Compared to the angular part, however, the radial equation is complicated and unconventional.
Though some exact results have been known \cite{leaver1985, leaver1986, mano1996}, the complexity often prevents us from understanding analytic properties.
There are attempts to transform the radial Teukolsky equation into more convenient forms \cite{chandrasekhar1976, sasaki1982}.
Unfortunately the resultant equations still look quite complicated even in the homogeneous case.

Surprisingly, there exists another description. We show our main result here. 
A new differential equation we found in this paper is the following:
\begin{equation}
\begin{aligned}
\left[ f(z)\frac{d}{dz}f(z) \frac{d}{dz} +(2M\omega)^2-V(z) \right] \phi(z)=0,
\end{aligned}
\label{eq:new}
\end{equation}
where $M$ is the mass of black hole, $\omega$ is the frequency and $f(z)=1-1/z$. 
We have introduced a rescaled radial variable $z$ so that $z=1$ and $z=\infty$ are the event horizon and the spacial infinity, respectively. See \eqref{eq:z-def}.
The potential is explicitly given by
\begin{equation}
\begin{aligned}
V(z)=f(z) \biggl[ 4c^2+\frac{4c(m-c)}{z}
+\frac{{}_sA_{\ell m}+s(s+1)-c(2m-c)}{z^2}-\frac{s^2-1}{z^3} \biggr],
\end{aligned}
\label{eq:V}
\end{equation}
where $c=a\omega$ is related to the rotation parameter $a$.
Other parameters in the potential are explained in the next section.
We conjecture that this simple equation is exactly \textit{isospectral} to the radial Teukolsky equation.
This is far from obvious. At the moment, we do not have a rigorous proof nor a direct derivation.\footnote{It seems that the equation \eqref{eq:new} can be derived by a highly non-trivial integral transformation in \cite{whiting1989, andersson2017}. We are grateful to Masato Nozawa for pointing out this fact with his great insight. We stress that the authors in \cite{whiting1989, andersson2017} did not mention the particularly useful equation \eqref{eq:new}.}

The equation \eqref{eq:new} has several advantages compared to the original radial Teukolsky equation (see \eqref{eq:radial}).
First, it reduces to the Regge--Wheeler equation smoothly in the non-rotating limit, while the Teukolsky equation does not.
Second, the coefficients in the potential \eqref{eq:V} are all real. Though the Kerr metric in the Boyer--Lindquist coordinates does not contain any imaginary numbers, the Teukolsky potential \eqref{eq:V_T} has unconventional imaginary coefficients. We expect that the equation \eqref{eq:new} will be useful in physical analyses.

We unexpectedly noticed this equivalence with the help of recent developments in theoretical high-energy physics and in mathematical physics. 
Once one accepts the isospectality, our equation is a powerful tool for analytical studies as well as for numerical ones.
We stress that the equation \eqref{eq:new} is the same form as master equations for spherically symmetric black holes if $c$ and ${}_sA_{\ell m}$ are regarded as given parameters.
To our knowledge, this equation has not been recognized in any references so far.

Let us illustrate a relationship to supersymmetric gauge theories.
It is well-known that both the angular and the radial Teukolsky equations are related to the confluent Heun equation (CHE).
Interestingly, the same differential equation appears in the $\cN=2$ SU(2) supersymmetric quantum chromodynamics (SQCD) with three fundamental matters.
Using this correspondence, Aminov, Grassi and the author recently proposed exact Bohr--Sommerfeld quantization conditions that determine quasinormal mode (QNM) frequencies of the Kerr black holes \cite{aminov2020}.

The idea we use in this paper is as follows.
Once moving to the SQCD, we have an obvious symmetry for flavor masses. The computation in \cite{aminov2020} strongly implies that the QNM spectrum respects it. Curiously, the same symmetry is \textit{not} manifest at the level of differential equations. As a consequence, an exchange of the masses leads to two apparently different equations.
In this way, we arrive at a dual counterpart of the Teukolsky equation.
Fortunately the resultant equation is simpler.

\section{The Teukolsky equation}
We start with the homogeneous (or vacuum) Teukolsky equation \cite{teukolsky1972} that governs perturbation of rotating black holes.
The Tuekolsky equation is a separable partial differential equation in Boyer-Lindquist coordinates.
After separation of variables, its angular part reads
\begin{equation}
\begin{aligned}
\biggl[ \frac{d}{dx}(1-x^2)\frac{d}{dx}+(cx)^2-2csx+{}_sA_{\ell m}+s
-\frac{(m+sx)^2}{1-x^2} \biggr] {}_sS_{\ell m}(x)=0,
\end{aligned}
\label{eq:angular}
\end{equation}
where $x=\cos \theta$ with the angular variable $\theta$, and $s$ is a spin-weight of a perturbing field. 
The eigenfunction ${}_sS_{\ell m}(x)$ is called the spin-weighted spheroidal harmonics in the literature.
The separation constant ${}_sA_{\ell m}$ is determined by the regularity condition at $x=\pm 1$ of ${}_sS_{\ell m}(x)$.
It reduces to $\l(\l+1)-s(s+1)$ in $c \to 0$.
Recently an analytic expression of ${}_sA_{\ell m}$ in terms of a known special function in the $\mathcal{N}=2$ SQCD was conjectured in \cite{aminov2020}.

The radial Teukolsky equation looks more unconventional:
\begin{equation}
\begin{aligned}
\Delta(r) R''(r)+(s+1)\Delta'(r) R'(r)+V_T(r) R(r)=0,
\end{aligned}
\label{eq:radial}
\end{equation}
where $\Delta(r)=r^2-2Mr+a^2$. The potential is
\begin{equation}
\begin{aligned}
V_T(r)=\frac{K(r)^2-2is(r-M)K(r)}{\Delta(r)}-{}_sA_{\ell m}
+4is\omega r+2am\omega-a^2\omega^2,
\end{aligned}
\label{eq:V_T}
\end{equation}
where $K(r)=(r^2+a^2)\omega-am$.
Note that the radial differential equation \eqref{eq:radial} has (regular) singular points at $r=r_\pm:=M\pm \sqrt{M^2-a^2}$.
In the next section, we map this radial wave equation into a much simpler form.

\section{A hidden symmetry}
The angular and the radial parts of the Teukolsky equation have the same singularity structure: two regular singular points and one irregular singular point. One can rewrite them as the normal form of the CHE \cite{aminov2020}.
The relation between the Teukolsky equation and the confluent Heun equation was discussed in detail in \cite{fiziev2009, fiziev2010}.
For the angular part, we change the variable $z=(1+x)/2$, and define $y(z):=\sqrt{1-x^2}{}_sS_{\ell m}(x)/2$.
Then we obtain
\begin{equation}
\begin{aligned}
y''(z)+Q(z) y(z)=0,
\end{aligned}
\label{eq:radial-2}
\end{equation}
where $Q(z)$ takes the form
\begin{equation}
\begin{aligned}
Q(z)=\frac{1}{z^2(z-1)^2} \sum_{j=0}^4 q_j z^j.
\end{aligned}
\label{eq:Q}
\end{equation}
The coefficients $q_j$ are computed straightforwardly.
Similarly, defining 
\begin{equation}
\begin{aligned}
z=\frac{r-r_-}{r_+-r_-}
\end{aligned}
\label{eq:z-def}
\end{equation}
and $y(z):=\Delta(r)^{(s+1)/2} R(r)$ for the radial part,
we obtain the same form as \eqref{eq:radial-2} and \eqref{eq:Q} with different coefficients.

The Teukolsky equation has an unexpected connection with the quantum Seiberg--Witten geometry \cite{seiberg1994, seiberg1994a, nekrasov2010} of the $\mathcal{N}=2$ SU(2) SQCD with three hypermultiplets \cite{fucito2011, zenkevich2011}. We compare \eqref{eq:radial-2} and \eqref{eq:Q} with eqs. (2.28) and (2.29) in \cite{aminov2020}.
The matching condition leads to eight possible identifications,
but it is sufficient to take one of them for our purpose due to a symmetric property of the CHE.
For the angular part, we have, for instance,
\begin{equation} 
\begin{aligned}
\Lambda_3&=16c,\quad E=-{}_sA_{\ell m}-s(s+1)-c^2-\frac{1}{4}, \\
m_1&=-m, \quad m_2=m_3=-s, \\
\end{aligned}
\label{eq:id0}
\end{equation}
where $\Lambda_3$ is the dynamical scale, $E$ is a moduli parameter, and $m_i$ are masses of the matters. 
For the radial part, we have
\begin{equation} 
\begin{aligned}
\Lambda_3&=16i\omega \sqrt{M^2-a^2}, \\
E&=-{}_sA_{\ell m}-s(s+1)+(8M^2-a^2)\omega^2-\frac{1}{4}, \\
m_1&=s+2iM\omega,\qquad m_3=-s+2iM\omega, \\
m_2&=\frac{i(2M^2\omega-am)}{\sqrt{M^2-a^2}}. \\
\end{aligned}
\label{eq:id}
\end{equation}
A bonus of these identifications is that we can use a symmetry in permutations of the three masses.
This symmetry is invisible in the Teukolsky equation itself.
All the physical quantities in the SQCD are invariant in these permutations.
In the differential equation, the symmetry in $m_1 \leftrightarrow m_2$ is manifest, but the others are not.
This curious fact motivates us to interchange the masses $m_2$ and $m_3$ in \eqref{eq:id}.
After this change, we go back to the above argument conversely.
We finally arrive at our proposal \eqref{eq:new}.
As was shown in \cite{aminov2020}, the QNM spectrum of the Teukolsky equation is determined by Nekrasov's function \cite{nekrasov2003} in the SQCD.
This function is of course invariant in the mass exchange. Therefore we conjecture that the radial Teukolsky equation \eqref{eq:radial} is exactly isospectral to \eqref{eq:new}.

In the extremal limit, things are more involved. The CHE \eqref{eq:radial-2} with \eqref{eq:id} becomes ill-defined due to the divergence of $m_2$. In this case, two regular singular points in the radial Teukolsky equation meet together, and the resulting equation reduces to the double confluent Heun equation (DCHE). 
Its singularity structure is obviously different from that in \eqref{eq:Q}.
Interestingly, the break down of \eqref{eq:radial-2} is avoided by interchange $m_2$ and $m_3$
because $m_3$ necessarily appears as $m_3 \Lambda_3$,
which is finite in the extremal limit.
As a result, the CHE \eqref{eq:radial-2} still holds, but its Poincar\'e rank at $z=\infty$ reduces from $1$ to $1/2$.
One might wonder that the equations \eqref{eq:new} and \eqref{eq:radial} lead to the two equations with different singularities: the CHE with rank $1/2$ and the DCHE.
However, solutions to these two equations are related by the Laplace transform \cite{ronveaux1995}.\footnote{We thank Hajime Nagoya for pointing out this fact.}
The similar result is also found on the gauge theory side.
The extremal limit just corresponds to a matter decoupling limit of the SQCD \cite{aminov2020}.
There are two equivalent realizations of the SQCD with two flavors \cite{gaiotto2013}.
The above two cases just correspond to these two realizations.
Interestingly, the similar structure was also found in the mode stability analysis \cite{teixeiradacosta2020}.
We claim that the equations \eqref{eq:new} and \eqref{eq:radial} are isospectral even in the extremal limit.

We play the same game for the angular equation. In this case, we rearrange $m_1=m_2=-s$ and $m_3=-m$, and 
obtain the following dual equation:
\begin{align}
\biggl[ \frac{d}{dx}(1-x^2)\frac{d}{dx}+(cx)^2&-2cmx+{}_sA_{\ell m}+s(s+1)
-\frac{2s^2}{1-x} \biggr] {}_s\widetilde{S}_{\l m}(x)=0.
\label{eq:angular-dual}
\end{align}
It looks more asymmetric than \eqref{eq:angular}. We do not find its particular advantage in this case because the angular Teukolsky equation \eqref{eq:angular} has a nice analogy with the scalar spheroidal harmonics or with the spin-weighted spherical harmonics.
A merit is rather that one can easily test the isospectrality of these two eigenvalue problems.
In fact, the regularity condition of ${}_s\widetilde{S}_{\l m}(x)$ at $x=\pm 1$ leads to the same spectrum of ${}_sA_{\ell m}$ as the angular Teukolsky equation \eqref{eq:angular}. This test directly supports our idea.

\section{Checking the isospectrality}\label{sec:test}
In this section, we show two direct quantitative tests of our isospectral conjecture.
Since we do not have its rigorous mathematical proof, these checks are significant.
The first one is to derive perturbative series of the QNM frequencies in the slowly rotating limit.
The second one is to solve the eigenvalue problem numerically.
These two computations are easily compared against the results for the Teukolsky equation.
We show a brief idea to accomplish them.

Let us first see boundary conditions. Our potential \eqref{eq:V} asymptotically behaves as $V \to 0$ ($z_* \to -\infty$) and $V \to 4c^2$ ($z_* \to \infty$),
where the tortoise variable $z_*$ is defined by $z_*=z+\log(z-1)$.
It also has a wall on the real axis of $z_*$. 
Therefore we impose a resonance boundary condition where only the waves getting away from the potential wall exist.
That is, we have $\phi \sim e^{-2iM\omega z_*}$ ($z_* \to -\infty$) and $\phi \sim e^{2i\mu \omega z_*}$ ($z_* \to \infty$) where $\mu=\sqrt{M^2-a^2}$.
In black hole perturbation theory, this boundary condition is nothing but the QNM boundary condition.
Of course, in the case of $a=0$, this is equivalent to the condition for the Schwarzschild black hole.

We compute slow spin corrections to the resonance frequency.
To do so, we use a systematic formalism proposed in \cite{hatsudaa}.
The potential \eqref{eq:V} has a natural deformation parameter.
It is regarded as a deformation of the spin-weighted Regge--Wheeler potential by the dimensionless parameter $c$.
To compute the perturbative corrections, we need the corrections to the angular eigenvalue:
\begin{equation}
\begin{aligned}
{}_sA_{\ell m}=\sum_{j=0}^\infty {}_sA_{\ell m}^{(j)} c^j.
\end{aligned}
\end{equation}
The explicit forms of ${}_sA_{\ell m}^{(j)}$ up to $j=6$ are found in \cite{berti2006, berti2006a}.
Also the analytic expressions to $j=12$ are conjectured in \cite{aminov2020}.
Unfortunately, for our purpose, we need higher-order information.
We apply the well-known perturbative method \textit{\`a la} Bender and Wu \cite{bender1969}.
This method allows us to compute the small-$c$ expansion of ${}_sA_{\ell m}$ to sufficiently high orders.

Using the formalism in \cite{hatsudaa}, we can easily compute the following perturbative corrections:
\begin{equation}
\begin{aligned}
M({}_s\omega_{\l m} -{}_s\omega_{\l}^{(0)})=\sum_{j=1}^\infty {}_sv_{\l m}^{(j)} c^{j},
\end{aligned}
\label{eq:correction}
\end{equation}
where ${}_s\omega_{\l}^{(0)}$ is the frequency for the Schwarzschild black holes.
However, we are rather interested in the perturbative series in the dimensionless rotation parameter $a/M$.
To do so, we plug \eqref{eq:correction} into $c=a\omega$, and inversely expand $c$ in terms of $a/M$.
From it, we immediately obtain the following series:
\begin{equation}
\begin{aligned}
M({}_s\omega_{\l m} -{}_s\omega_{\l}^{(0)})=\sum_{j=1}^\infty {}_sw_{\l m}^{(j)} \Bigl( \frac{a}{M} \Bigr)^{j}.
\end{aligned}
\label{eq:correction-2}
\end{equation}
Let us consider symmetries of this perturbative series. The perturbative coefficients of ${}_sA_{\ell m}$ have a symmetric relation ${}_sA_{\ell, -m}^{(j)}=(-1)^j {}_sA_{\ell m}^{(j)}$. This symmetry is understood as a consequence of an invariance under $(x, c, m) \to (-x, -c, -m)$ in the angular equation \eqref{eq:angular}. Also the potential \eqref{eq:V} is invariant under $(c,m) \to (-c,-m)$. Therefore we conclude that the dimensionless coefficients have the following symmetric relation:
\begin{equation}
\begin{aligned}
{}_sw_{\l, -m}^{(j)}=(-1)^j {}_sw_{\l m}^{(j)}.
\end{aligned}
\label{eq:sym}
\end{equation}
In particular, there are no odd order corrections for $m=0$: ${}_sw_{\l 0}^{(2p+1)}=0$ ($\forall p \in \mathbb{Z}$).
In Table~\ref{tab:pert-QNM}, we show the first twelve perturbative coefficients to the fundamental QNM frequency for $(s,\l)=(-2,2)$ obtained in this way.
The first and second order corrections are perfectly consistent with a recent computation in a different approach \cite{hatsuda2020b}.

\begin{table}[t]
\caption{The first twelve non-zero perturbative coefficients in \eqref{eq:correction-2} to the lowest overtone frequency for $(s,\l)=(-2,2)$. The coefficients for $m=-1,-2$ are known from the symmetry \eqref{eq:sym}.}

\begin{center}
\begin{tabular}{c|rr|rr|rr}
\hline
$j$ & $\real ({}_{-2}w_{20}^{(2j)})$ & $\im ({}_{-2}w_{20}^{(2j)})$ & $\real ({}_{-2}w_{21}^{(j)})$ & $\im ({}_{-2}w_{21}^{(j)})$ & $\real ({}_{-2}w_{22}^{(j)})$ & $\im ({}_{-2}w_{22}^{(j)})$  \\
\hline
$1$ & $0.03591312868$ & $0.00638178925$  & $0.062883080$ & $0.000997935$ & $0.125766159$ & $0.001995870$\\
$2$ & $0.00968816868$ & $0.00405143823$ & $0.044869919$ & $0.006090568$ & $0.071740290$ & $0.005216904$ \\
$3$ & $0.00359183682$ & $0.00240859190$ & $0.021816430$ & $0.002879144$ & $0.048026343$ & $0.004409416$\\
$4$ & $0.00152055523$ & $0.00149553496$ & $0.016285342$ & $0.004080204$ & $0.035129263$ & $0.004100755$\\
$5$ & $0.00067437679$ & $0.00096228201$ & $0.010938644$ & $0.002546043$ & $0.027212758$ & $0.003743113$\\
$6$ & $0.00029193511$ & $0.00063476253$ & $0.008383298$ & $0.002756152$ & $0.021794044$ & $0.003388413$\\
$7$ & $0.00010966322$ & $0.00042559175$ & $0.006433797$ & $0.002082095$ & $0.017927392$ & $0.003085350$\\
$8$ & $0.00002139324$ & $0.00028803117$ & $0.005083673$ & $0.002014605$ & $0.015053622$ & $0.002783931$\\
$9$ & $-0.0000202845$ & $0.00019557831$ & $0.004157778$ & $0.001684674$ & $0.012844972$ & $0.002513832$\\
$10$ & $-0.0000381452$ & $0.00013245712$ & $0.003387223$ & $0.001556853$ & $0.011120330$ & $0.002267731$\\
$11$ & $-0.0000437166$ & $0.00008889498$ & $0.002864656$ & $0.001373315$ & $0.009739945$ & $0.002045238$\\
$12$ & $-0.0000430492$ & $0.00005863675$ & $0.002398475$ & $0.001250179$ & $0.008621821$ & $0.001849651$\\
\hline
\end{tabular}
\end{center}
\label{tab:pert-QNM}
\end{table}%

\begin{figure}[tb]
\centering
 \includegraphics[width=0.6\linewidth]{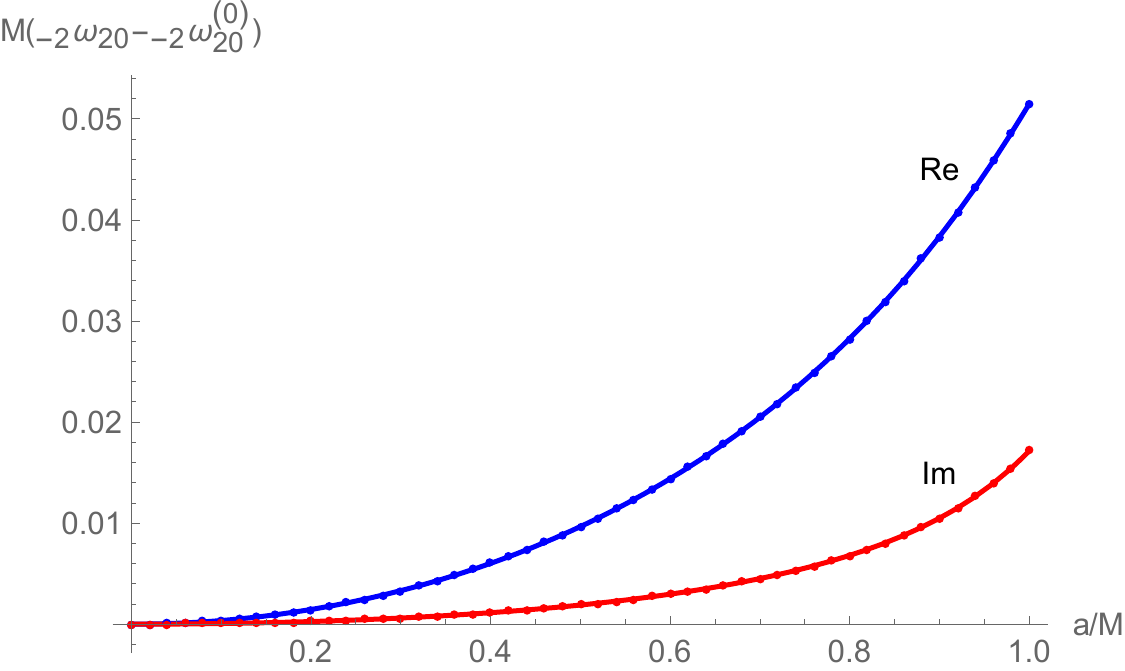}
\caption{The $[12/12]$ Pad\'e approximant (the solid lines) of \eqref{eq:correction-2} for $(s,\l,m)=(-2,2,0)$ is compared with the numerical data (the points) from the Teukolsky equation. The deviations from the numerical values at the extremal case $a/M=1$ are about 0.004\% for the real part and about 0.01\% for the imaginary part.}
\label{fig:m0}
\end{figure}

\begin{figure}[tb]
  \begin{minipage}[b]{0.5\linewidth}
    \centering
    \includegraphics[width=0.95\linewidth]{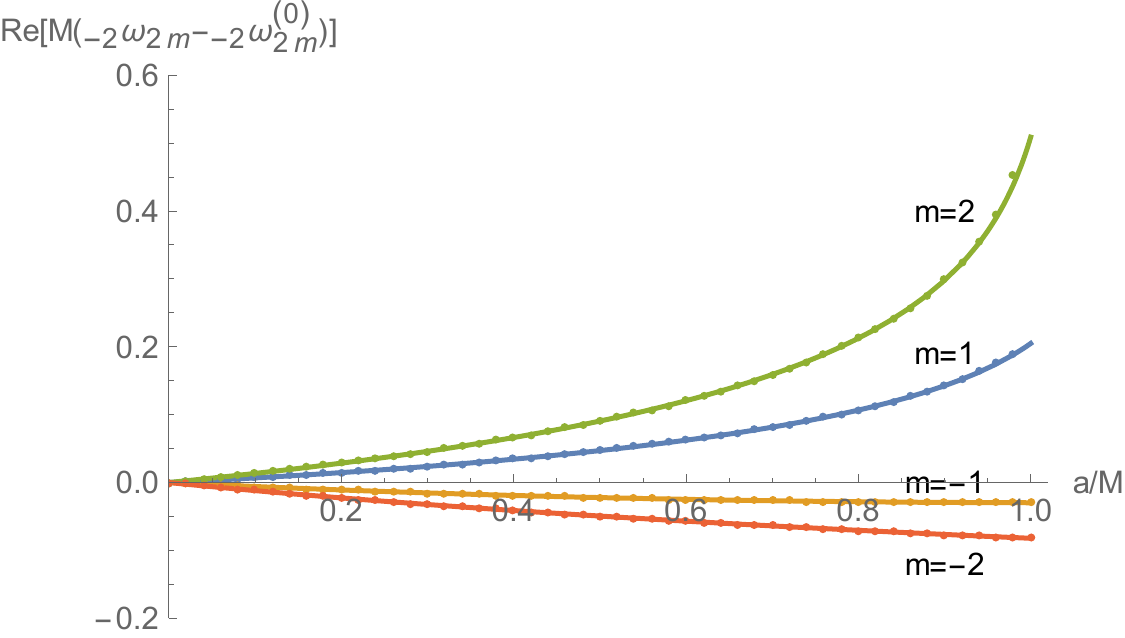}
  \end{minipage}
    \begin{minipage}[b]{0.5\linewidth}
    \centering
    \includegraphics[width=0.95\linewidth]{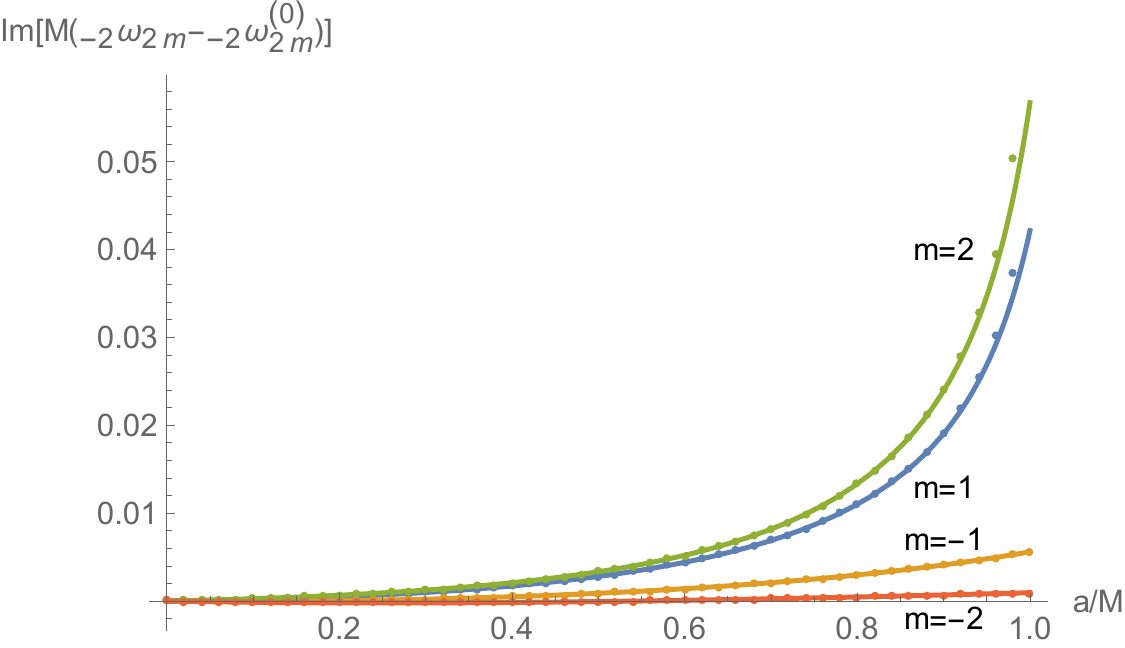}
  \end{minipage}
  \caption{The $[6/6]$ Pad\'e approximants (the solid lines) of \eqref{eq:correction-2} for $m \ne 0$ with $(s,\l)=(-2,2)$ are compared with the numerical data (the points) from the Teukolsky equation. We separately show the real part in the upper panel and the imaginary part in the lower panel. }
  \label{fig:mnot0}
\end{figure}

Now we compare these results with numerical data.
Figure \ref{fig:m0} shows the fundamental QNM frequency for $(s,\ell, m)=(-2,2,0)$.
We have used publicly available numerical data obtained by the Teukolsky equation \cite{berti2006b, berti2009, zotero-589, richartz2016}, which are shown by the points.
We compare these data with a rational approximation constructed from the perturbative data. We compute the diagonal Pad\'e approximant of degree 12 by using the 24th order data in Table~\ref{tab:pert-QNM}.
The Pad\'e approximant explains the numerical data in the whole region $0 \leq a/M \leq 1$.
In Figure \ref{fig:mnot0}, we also show the fundamental frequencies for $m \ne 0$ with $(s,\l)=(-2,2)$.
In all cases, the Pad\'e approximants can be extrapolated to $a/M\sim 0.9$.
For instance, the most deviated value at $a/M=0.9$ is for the imaginary part with $m=2$. Even in this worst case, the deviation of the $[6/6]$ Pad\'e approximant from the numerical value is about $0.75\%$.

For a more quantitative check, we solve the eigenvalue problem numerically.
In the angular problem, many numerical techniques have been known.
Here we use a great \textit{Mathematica} package on the spin-weighted spheroidal harmonics \cite{BHPT} that evaluates ${}_sA_{\l m}$ for arbitrary complex $c$ quickly and precisely.

We are particularly interested in the radial problem.
To solve the radial eigenvalue problem \eqref{eq:new}, we use Leaver's continued fraction method \cite{leaver1985}.
We first change the variable by $\zeta=1-1/z$. The two singular points $z=1,\infty$ are mapped to $\zeta=0,1$ respectively.
Taking into account the asymptotic solutions at $\zeta=0,1$, we put the following ansatz of the eigenfunction:
\begin{equation}
\begin{aligned}
\phi=\zeta^{-2iM\omega}(1-\zeta)^{\frac{2M^2\omega-am}{i\mu}} e^{\frac{2i\mu\omega}{1-\zeta}} \sum_{n=0}^\infty a_n \zeta^n.
\end{aligned}
\label{eq:ansatz}
\end{equation}
The coefficients satisfy a three-term recurrence relation $\alpha_n a_{n+1}+\beta_{n} a_n+\gamma_n a_{n-1}=0$ where
\begin{align}
\alpha_n&=\mu^2(n+1)(n+1-4iM\omega), \notag\\
\beta_n&=\mu(2r_+\omega+in)[(2r_+(r_++M)\omega-2am+i\mu(n+2)]\notag\\
&\quad -ia\mu(m-2a\omega+2i\mu m\omega)\\
&\quad -\mu^2(n^2+{}_sA_{\l m}+s+1+4r_+^2\omega^2+a^2\omega^2), \notag\\
\gamma_n&=-(2Mr_+\omega-am+i\mu n)^2-\mu^2s^2.\notag
\end{align}
Recall that $\mu=\sqrt{M^2-a^2}$ and $r_+=M+\mu$. 
The boundary condition at $\zeta=1$ finally requires the so-called continued fraction equation:
\begin{equation}
\begin{aligned}
\beta_0=\frac{\alpha_0 \gamma_1}{\beta_1-}\frac{\alpha_1\gamma_2}{\beta_2-}\frac{\alpha_2 \gamma_3}{\beta_3-}\cdots.
\end{aligned}
\end{equation}
We have followed the notation in \cite{leaver1985}.
This equation determines the QNM frequency $\omega$.
Of course the equation here looks quite different from Leaver's original result for the radial Teukolsky equation.
We compare numerical values obtained in this way against the known results in the literature.
We find good agreement with \cite{leaver1985, zotero-589} within numerical errors.
This is a highly non-trivial test of the isospectrality.

In the extremal limit, since the Poincar\'e rank at the irregular singular point reduces to $1/2$, we need to modify the ansatz \eqref{eq:ansatz} slightly.
This modification leads to a four-term relation, and the resulting continued fraction equation correctly reproduces the result in \cite{richartz2016}.

\section{Concluding remarks}\label{sec:conclusion}
In this paper, starting with the radial Teukolsky equation, we found an isospectral differential equation.
In the gauge theoretical perspective,
it is clear why the Teukolsky equation is complicated.
This is caused by an asymmetric mass choice. A more symmetric choice leads to a beautiful equation.
Our result is particularly useful to analyze the QNM frequency of the Kerr black holes.
In fact, we derived for the first time the high-order perturbative corrections to the QNM frequency for the slowly rotating Kerr black holes.
We also performed numerical tests that strongly support our conjecture.

We point out that there is another third order differential equation in the three-flavored SQCD \cite{fucito2011, zenkevich2011}. Probably this equation is also isospectral to the Teukolsky equation. In this third order equation, the mass symmetry is manifest. It is interesting to study it in detail.


The isospectrality is mathematically interesting in its own right. For instance, the similar isospectrality was recently applied for studying the mode stability of the Kerr-de Sitter black holes \cite{casals2021}. However, we still have many open problems.
One of the most mysterious questions is a physical origin of the differential equation \eqref{eq:new}.
Can we derive it directly in black hole perturbation theory?
Or can we find a concrete relation between Teukolsky's radial eigenfunction $R(r)$ and our eigenfunction $\phi(z)$?
For more physical applications, we need the inhomogeneous equation including sources.
It is very important to clarify them.
It is also interesting to extend our result to higher dimensional or asymptotically non-flat cases.

Similarly, relations to the equations in \cite{chandrasekhar1976, sasaki1982} are also unclear.
As well as ours, these equations also reduce to the Regge--Wheeler equation in the non-rotating limit.
There should be a relation. It is nice to find it.

It is known that near the extremal limit the QNM spectrum shows a transition between two branches called damped modes and zero-damping modes \cite{yang2013}. 
Interestingly, the second mass in \eqref{eq:id} does not diverge in $a \to M$ for the zero-damping modes.
We do not understand its physical interpretation.

The potential \eqref{eq:V} with $s=-2$ simply reduces to the Regge--Wheeler potential in the limit $c \to 0$.
The Regge--Wheeler potential is obtained by the odd-parity gravitational perturbation, while the even-parity perturbation
leads to the Zerilli potential. It is unclear why our dual equation is a deformation of the Regge--Wheeler potential
rather than of the Zerilli potential.
Is there a further equivalent description based on the Zerilli potential?

\section*{Acknowledgments}
We thank Gleb Aminov, Alba Grassi, Katsushi Ito, Masashi Kimura, Hajime Nagoya and Masato Nozawa for valuable discussions.
This work is supported by JSPS KAKENHI Grant Number JP18K03657.

\bibliographystyle{amsmod}
\bibliography{Alt-Teu}

\end{document}